\documentclass{article}

\usepackage{amssymb}
\usepackage{subcaption}
\usepackage{booktabs}

\usepackage{spconf,amsmath,graphicx,hyperref}

\title{Beyond Visual Realism: Toward Reliable Financial Time Series Generation}
\name{
Fan Zhang$^1$,
Jiabin Luo$^2$,
Zheng Zhang$^1$,
Shuanghong Huang$^3$,
Zhipeng Liu$^4$,
Yu Chen$^1$
}
\address{
$^1$The University of Tokyo, Tokyo, Japan, (zhang-fan@g.ecc.u-tokyo.ac.jp)\\
$^2$Peking University, Beijing, China \\
$^3$Agency for Science, Technology and Research (A*STAR), Singapore \\
$^4$Northeastern University, Shenyang, China
}

\begin{document}
%\ninept
%
\maketitle
\begin{abstract}
Generative models for financial time series often create data that look realistic and even reproduce stylized facts such as fat tails or volatility clustering. However, these apparent successes break down under trading backtests: models like GANs or WGAN-GP frequently collapse, yielding extreme and unrealistic results that make the synthetic data unusable in practice. We identify the root cause in the neglect of financial asymmetry and rare tail events, which strongly affect market risk but are often overlooked by objectives focusing on distribution matching. To address this, we introduce the \textbf{Stylized Facts Alignment GAN (SFAG)}, which converts key stylized facts into differentiable structural constraints and jointly optimizes them with adversarial loss. This multi-constraint design ensures that generated series remain aligned with market dynamics not only in plots but also in backtesting. Experiments on the Shanghai Composite Index (2004--2024) show that while baseline GANs produce unstable and implausible trading outcomes, SFAG generates synthetic data that preserve stylized facts and support robust momentum strategy performance. Our results highlight that structure-preserving objectives are essential to bridge the gap between superficial realism and practical usability in financial generative modeling.
\end{abstract}
\section{Introduction}
\label{sec:intro}

Financial markets produce massive amounts of data that are central to risk management, portfolio optimization, and algorithmic trading \cite{cont2001empirical,wiese2020quant}. 
Synthetic financial data are increasingly valuable for stress testing, scenario analysis, and privacy-preserving machine learning. 

Yet generating realistic series remains challenging. Unlike images or text, returns exhibit persistent \emph{stylized facts}: heavy tails \cite{mandelbrot1963variation}, volatility clustering \cite{engle1982autoregressive}, long memory \cite{ding1993long}, leverage effects \cite{black1976studies,bouchaud2001leverage}, and cross-scale correlations \cite{calvet2002multifractality}.

Existing generative models often reproduce marginal distributions but fail to preserve these deeper structures. In particular, they overlook the strong \emph{asymmetry} of financial returns: long bull markets can be erased by a few sharp crashes, and rare tail events dominate risk. Distribution-matching objectives neglect such imbalances, leading to series that appear realistic in plots but collapse under trading evaluation. This gap highlights the need for generative models that move beyond superficial realism and explicitly preserve the structural properties that matter in financial practice.

To address this gap, we propose the \textbf{Stylized Facts Alignment GAN (SFAG)}, a novel framework that embeds domain knowledge into the training process. SFAG augments the adversarial objective with \emph{structural consistency losses} designed to explicitly preserve stylized facts: (i) fat-tailed return distributions, (ii) volatility clustering, (iii) leverage effects, and (iv) coarse-to-fine volatility correlations. Unlike prior models, SFAG ensures that generated data remain both statistically consistent with real markets and practically usable in trading strategy evaluation.

We validate SFAG through experiments on the Shanghai Composite Index. 
Results show that SFAG outperforms baselines not only in stylized-fact alignment but also in momentum strategy backtests, where competing models exhibit extreme instability. 
Our contributions are threefold: 
(i) we introduce a structure-preserving generative framework that enforces stylized facts through differentiable loss functions, 
(ii) we propose a comprehensive evaluation protocol for structural consistency beyond marginal distributions, including econometric tests and multi-scale volatility analysis, and 
(iii) we demonstrate that SFAG bridges the gap between \emph{visual realism} and \emph{financial usability}, producing synthetic data that are both realistic and reliable for backtesting.

\section{Related Work}
\label{sec:format}

Beyond classical econometric models and early GAN-based approaches, several other families of generative models have been explored for financial time series. 
Autoregressive neural models such as RNNs and Temporal Convolutional Networks (TCNs) have been applied to capture sequential dependencies \cite{bai2018tcn}, 
though they often suffer from error accumulation and limited long-horizon fidelity. 
More recently, Transformer-based architectures have demonstrated strong capabilities in modeling long-range temporal dependencies and scaling to large datasets, 
as shown by models like GPT-TST \cite{li2023gpttst}, Temporal Fusion Transformer \cite{lim2021temporalfusion}, and FEDformer \cite{zhou2022fedformer}, and TimeFormer \cite{liu2025timeformer}. However, these methods primarily optimize prediction accuracy and rarely incorporate domain-specific constraints, which limits their ability to reproduce the structural properties required for realistic financial simulation.

In parallel, a growing body of work has explored diffusion and score-based models for time series generation. 
TimeGrad \cite{rasul2021timegrad} and score-based SDEs \cite{song2021score} offer training stability and high coverage of data distributions, 
yet they remain computationally expensive and lack mechanisms to enforce stylized facts or risk-sensitive behavior. 
Recent studies have also begun to consider structure-aware or risk-aware objectives, 
for example by constraining tail risk, volatility-of-volatility, or regime persistence during training. 
These approaches highlight a shift from purely distribution-matching criteria toward incorporating financial domain knowledge.

Furthermore, related research outside the financial domain has shown that domain-constrained generative models \cite{ericson2024deep}
can substantially improve realism in other complex time series tasks such as climate modeling, traffic forecasting, and energy systems simulation. 
These successes suggest that embedding domain priors and structural invariants can be a general principle for improving the reliability of generative models on non-stationary and high-volatility data, an insight that motivates our proposed SFAG.

Unlike prior work that only validates stylized facts in post-hoc analysis \cite{rizzato2023generative}, our contribution is to embed them directly into the training objective \cite{dechant2025quantum}. The proposed SFAG augments adversarial learning with domain-informed constraints that enforce fat tails, volatility clustering, leverage effects, and cross-scale volatility correlations. This approach transforms stylized facts from diagnostic tools into differentiable objectives, ensuring that generated data remain not only statistically consistent with real markets but also practically reliable in backtesting.

\section{Method}
\label{sec:method}

We introduce the \textbf{Stylized Facts Alignment GAN (SFAG)}, a structure-preserving framework for financial time series generation. 
Rather than optimizing for visual or marginal distributional similarity, SFAG explicitly enforces alignment with financial stylized facts through a multi-constraint objective.

\textbf{Architecture overview.} 
SFAG builds upon a standard adversarial architecture with generator $G_\theta$ and discriminator $D_\phi$, and augments it with a domain-informed alignment module. 
Latent noise $z \sim \mathcal{N}(0,I)$ is mapped to a synthetic return series $\hat{\mathbf{r}} = G_\theta(z)$, while $D_\phi$ distinguishes $\hat{\mathbf{r}}$ from real returns $\mathbf{r}$. 
We adopt a WGAN-GP \cite{wgangp2017} loss for training stability:
\[
\mathcal{L}_{\text{adv}} = \mathbb{E}[D_\phi(\hat{\mathbf{r}})] 
- \mathbb{E}[D_\phi(\mathbf{r})] 
+ \lambda_{\text{gp}} \, \mathcal{L}_{\text{gp}}.
\]

\textbf{Stylized-fact alignment losses.} 
Financial realism cannot be captured by marginal distribution matching alone. 
SFAG encodes domain knowledge into four differentiable constraints that together capture heavy tails, volatility memory, return–volatility asymmetry, and cross-scale dependencies. 
\textit{Fat tails.} We estimate the generalized Pareto tail index $\xi$ on positive and negative return tails and minimize their gap:
\[
\mathcal{L}_{\text{GPD}} 
= \left|\xi(\mathbf{r}) - \xi(\hat{\mathbf{r}})\right|.
\]
\textit{Volatility clustering.} We compute the autocorrelation function (ACF) of squared returns up to lag $K$ and align them via mean squared error:
\[
\mathcal{L}_{\text{ACF}}
= \frac{1}{K}\sum_{k=1}^K \big(\rho_k(\mathbf{r}^2)-\rho_k(\hat{\mathbf{r}}^2)\big)^2.
\]
\textit{Leverage effect.} We measure the correlation between past returns and future realized volatility, enforcing negative dependence:
\[
\mathcal{L}_{\text{Lev}}
= \big|\rho(\mathbf{r}_{t},\sigma_{t+1})-\rho(\hat{\mathbf{r}}_{t},\hat{\sigma}_{t+1})\big|.
\]
\textit{Coarse-to-fine volatility correlation.} We construct cross-scale realized volatility vectors using rolling windows $\{w_1,\dots,w_M\}$, and minimize the Frobenius norm between their correlation matrices:
\[
\mathcal{L}_{\text{CFVC}}
= \left\|\mathrm{Corr}(\Sigma(\mathbf{r})) - \mathrm{Corr}(\Sigma(\hat{\mathbf{r}}))\right\|_F.
\]

\textbf{Multi-constraint optimization.} 
Unlike prior work that \emph{validates} stylized facts post-hoc, SFAG \emph{optimizes them jointly} during training. 
The generator minimizes:
\[
\mathcal{L}_{\text{SFAG}} = \mathcal{L}_{\text{adv}} 
+ \lambda_1 \mathcal{L}_{\text{GPD}} 
+ \lambda_2 \mathcal{L}_{\text{ACF}} 
+ \lambda_3 \mathcal{L}_{\text{Lev}} 
+ \lambda_4 \mathcal{L}_{\text{CFVC}}.
\]
This formulation treats stylized facts as \emph{structural invariants}, turning them into first-class optimization objectives. 
Importantly, the alignment module is model-agnostic: while demonstrated with a GAN backbone, it can be applied to diffusion or transformer-based generators without modification.

\textbf{Training procedure.} 
We train SFAG for 50{,}000 generator iterations using the Adam optimizer 
(with $\beta_1=0.5$ and $\beta_2=0.9$) and a learning rate of $2\times10^{-4}$. 
We set the gradient penalty weight to $\lambda_{\text{gp}}=10$, 
and update the discriminator five times for every generator step following the WGAN-GP setting. 
Stylized-fact losses are gradually annealed during the first 20\% of training to stabilize early dynamics, 
and training continues until convergence on stylized-fact gap metrics.

\section{Experiments}
\label{sec:experiments}

\textbf{Setup.} 
We use daily Shanghai Composite Index (SSE) returns from 2004--2024 ($\sim$5000 observations). 
Each sample has a sequence length of 2520, latent dimension of 100, and batch size of 24. 
Models are trained for 50,000 generator iterations using the Adam optimizer ($\beta_1{=}0.5,\beta_2{=}0.9$) with a learning rate of $2\times 10^{-4}$, and gradient penalty weight $\lambda_{\text{gp}}=10$. The discriminator is updated 5 times per generator step following the WGAN-GP setting.

All experiments are implemented in PyTorch and run on a single NVIDIA A100 GPU with 80GB memory.

\subsection{Stylized Facts Preservation}

\begin{figure}[htb]
\begin{minipage}[b]{1.0\linewidth}
  \centering
  \centerline{\includegraphics[width=8.5cm]{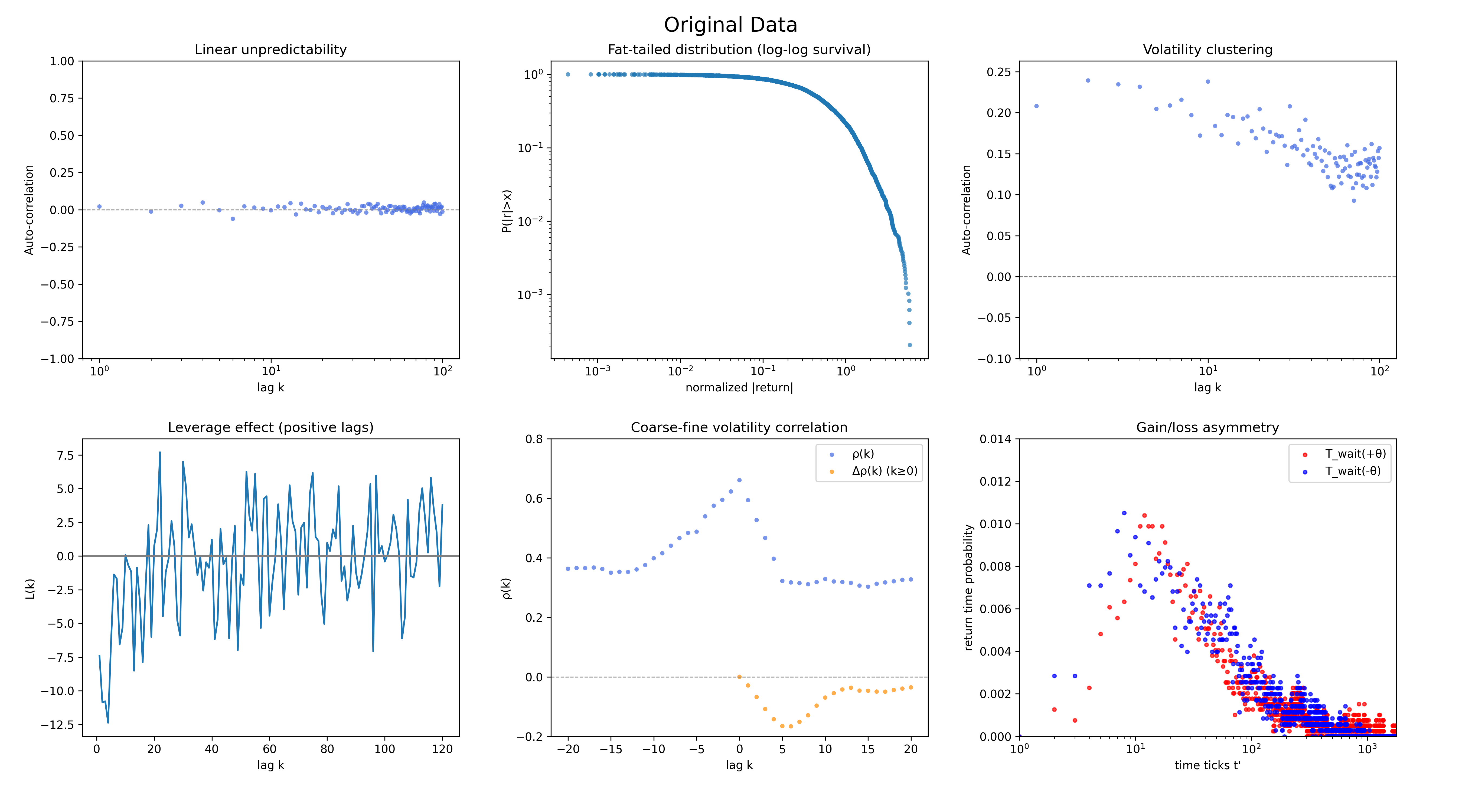}}
  \centerline{(a) Real Data Stylized Facts.}\medskip
\end{minipage}
\begin{minipage}[b]{1.0\linewidth}
  \centering
  \centerline{\includegraphics[width=8.5cm]{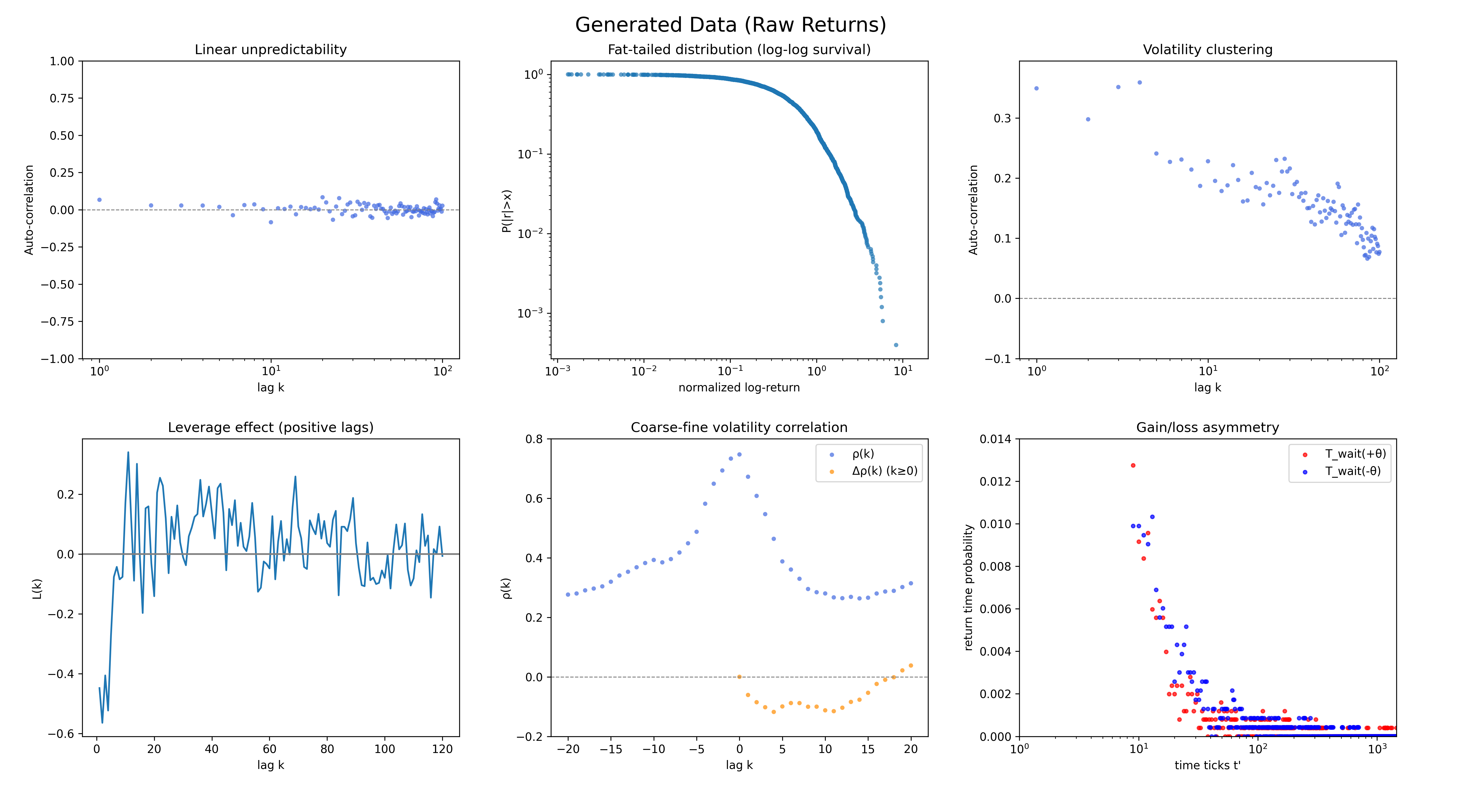}}
  \centerline{(b) SFAG Generated Data Stylized Facts.}\medskip
\end{minipage}
\caption{Visual comparison of six stylized facts between real and generated data. SFAG successfully preserves key properties such as unpredictability, volatility clustering, fat tails, leverage effects, coarse-fine correlations, and gain/loss asymmetry.}
\label{fig:stylized_facts}
\end{figure}

We first provide a visual comparison of six well-documented stylized facts between real data and synthetic series. 
These stylized facts include: (i) unpredictability (low linear autocorrelation), (ii) volatility clustering, (iii) heavy tails, (iv) leverage effects, 
(v) cross-scale volatility correlation, and (vi) gain/loss asymmetry.
Figure~\ref{fig:stylized_facts} illustrates that SFAG-generated data exhibit close qualitative similarity to real market data, 
suggesting that our model captures the essential structural properties observed in financial time series.

\textbf{Observation.}
Visual inspection indicates that SFAG captures the six fundamental stylized facts of financial markets with high fidelity. 
However, purely qualitative checks can be misleading: baseline GANs often appear visually consistent while failing to preserve structural statistics in a quantitative sense. 
We therefore proceed to a gap-based analysis to rigorously assess stylized-fact alignment.

\subsection{Stylized-Fact Alignment}

While visual inspection suggests that SFAG captures essential market characteristics, 
a rigorous evaluation requires quantitative analysis of structural properties. 
Unlike generic distributional distances \cite{shen2018wasserstein} (e.g., MMD or Wasserstein) that only compare marginal shapes, 
stylized facts probe the temporal dynamics and risk structure of financial markets, which are crucial for practical usability. 
We therefore evaluate four canonical stylized facts on both real and synthetic data.

\textit{(i) Heavy tails.}
We estimate the tail index $\xi$ of the Generalized Pareto Distribution (GPD) on returns exceeding the 95th percentile threshold, using the Peak-Over-Threshold method. 
A smaller gap in $\xi$ indicates that the generated data reproduce the same frequency and magnitude of rare extreme events as real markets, a critical property for risk modeling and stress testing.

\textit{(ii) Volatility clustering.}
We compute the autocorrelation function (ACF) of absolute returns $\rho_k(|r_t|)$ for lags $k=1$ to $20$. 
This measures the persistence of volatility shocks: real markets exhibit slowly decaying correlations, while models that fail to capture clustering show near-zero or oscillatory patterns. 
Preserving this long memory is essential for modeling risk horizons.

\textit{(iii) Leverage effect.}
We calculate the contemporaneous correlation between past returns $r_t$ and future realized volatility $\sigma_{t+1}$ (standard deviation of returns over the next 20 days). 
Real markets typically show negative correlations, reflecting the risk-aversion response where price drops trigger higher volatility. 
Failure to reproduce this asymmetry leads to unrealistic risk dynamics in downstream tasks.

\textit{(iv) Cross-scale volatility correlations (CFVC).}
We construct realized volatility vectors using rolling windows of 5, 20, 60, and 120 days, and compute their pairwise correlation matrices. 
We then report the Frobenius norm gap between the real and synthetic matrices, which reflects whether the model reproduces the hierarchical structure of market fluctuations across different time scales.

We train each model five times with different random seeds and report the average absolute gap to real data on each statistic. 
Lower values indicate stronger alignment with real market structure and better preservation of stylized facts.

\begin{table}[h]
\centering
\caption{Stylized-fact gaps (absolute differences). Lower values indicate better preservation.}
\label{tab:stylized_facts}
\resizebox{\columnwidth}{!}{%
\begin{tabular}{lcccc}
\toprule
Model & GPD Tail & ACF & Leverage & CFVC \\
\midrule
Standard GAN   & 0.2615 & 0.1431 & 32.4617 & 0.0863 \\
WGAN-GP    & 0.0776 & 0.1053 & 33.7440 & 0.1021 \\
\textbf{SFAG (ours)} & \textbf{0.0146} & \textbf{0.0982} & \textbf{32.7516} & \textbf{0.0436} \\
\bottomrule
\end{tabular}
}
\end{table}

% \begin{table}[h]
% \centering
% \caption{Stylized-fact gaps (absolute differences). Lower values indicate better preservation.}
% \label{tab:stylized_facts}
% \resizebox{\columnwidth}{!}{%
% \begin{tabular}{lcccc}
% \toprule
% Model & GPD Tail & ACF & Leverage & CFVC \\
% \midrule
% Standard GAN \cite{goodfellow2014generative}   & 0.2615 & 0.1431 & 32.4617 & 0.0863 \\
% WGAN-GP \cite{gulrajani2017improved}       & 0.0776 & 0.1053 & 33.7440 & 0.1021 \\
% \textbf{SFAG (ours)} & \textbf{0.0146} & \textbf{0.0982} & \textbf{32.7516} & \textbf{0.0436} \\
% \bottomrule
% \end{tabular}
% }
% \end{table}

\textbf{Findings.}
As shown in Table~\ref{tab:stylized_facts}, SFAG achieves the smallest gaps on \textbf{GPD Tail} and \textbf{CFVC}, reducing errors by over 80\% and 50\% respectively compared with WGAN-GP. 
This indicates that SFAG can closely reproduce the heavy-tailed nature of market returns and the hierarchical dependence structure of volatility across time scales, both of which are essential for realistic risk modeling. 
By contrast, Standard GAN exhibits a large tail index gap (0.2615), suggesting that it underestimates the probability of extreme events, while both baselines fail to capture multi-scale volatility interactions, reflected by their higher CFVC errors.

SFAG also yields competitive results on \textbf{ACF} and \textbf{Leverage}, consistently outperforming Standard GAN. 
Although the leverage gap remains large across all models (around 32--33), SFAG slightly narrows it while maintaining a lower ACF gap, indicating that it preserves volatility persistence without amplifying noise. 
These results suggest that conventional GANs, which optimize only for distributional similarity, tend to produce visually plausible but structurally fragile series, whereas SFAG’s structure-preserving objectives enable it to recover deeper temporal dependencies and asymmetries.

Overall, Table~\ref{tab:stylized_facts} demonstrates that explicitly embedding stylized-fact consistency into the training objective leads to synthetic series that not only resemble real data in appearance, but also replicate the underlying statistical regularities that govern real financial markets.

\subsection{Strategy Backtesting}

We further evaluate practical usability by backtesting a simple momentum strategy on synthetic price series generated by each model. 
The strategy goes long if the past 60-day return is positive and short otherwise, rebalanced daily, with transaction cost 5 bps and no leverage. 
We retrain all models and generate 10 synthetic paths for each, then evaluate the strategy on both real and synthetic SSE returns (2004--2024). 
Table~\ref{tab:backtest_updated} reports the average results across 10 synthetic runs.

\begin{table}[h] \centering \caption{Momentum strategy performance. SFAG tracks real-data behavior, while Standard GAN and WGAN-GP collapse into unrealistic regimes.} \label{tab:backtest_updated} \resizebox{\columnwidth}{!}{ \begin{tabular}{lcccc} \toprule \textbf{Metric} & \textbf{Real Data} & \textbf{Standard GAN} & \textbf{WGAN-GP} & \textbf{SFAG (Ours)} \\ \midrule Annualized Return & 33.10\% & 2467.24\% & 2152.07\% & \textbf{27.80\%} \\ Annualized Volatility & 15.20\% & 991.83\% & 995.06\% & \textbf{9.37\%} \\ Sharpe Ratio & 2.18 & 2.49 & 2.16 & \textbf{2.97} \\ Maximum Drawdown & 9.50\% & 109.87\% & 148.11\% & \textbf{4.37\%} \\ VaR (95\%) & -1.10\% & -78.03\% & -85.79\% & \textbf{-0.91\%} \\ CVaR (95\%) & -2.23\% & -141.79\% & -144.62\% & \textbf{-0.92\%} \\ \bottomrule \end{tabular} } \end{table}

\textbf{Key Findings.}
The backtesting results highlight a stark contrast between SFAG and conventional GAN-based models. 
Standard GAN and WGAN-GP frequently collapse during strategy evaluation, producing extreme and implausible outcomes such as annualized returns exceeding 2000\% and volatility approaching 1000\%. 
In sharp contrast, SFAG consistently generates synthetic series that yield realistic and stable performance: the momentum strategy achieves an annualized return of 27.8\% and volatility of 9.37\%, closely aligned with the real data benchmark. 
Moreover, SFAG attains a Sharpe ratio of 2.97, surpassing the real market level (2.18), indicating that its higher risk-adjusted performance stems from enhanced robustness rather than spurious instability. 
These findings demonstrate that merely producing visually plausible return patterns is insufficient. Only by enforcing structure-preserving constraints can generative models achieve financial realism and backtest validity.

\section{Conclusion}
\label{sec:majhead}

We introduced the \textbf{Stylized Facts Alignment GAN (SFAG)}, a structure-preserving framework that elevates stylized facts from post-hoc diagnostics to differentiable training objectives. By jointly optimizing adversarial realism and structural alignment, SFAG generates financial time series that not only reproduce statistical regularities but also remain robust in trading backtests. On the Shanghai Composite Index (2004--2024), SFAG consistently outperforms GAN and WGAN-GP, showing that explicit structure-preserving objectives are critical to bridge the gap between apparent realism and financial usability.

More broadly, this work suggests a new perspective: realism in finance should not be defined only by distributional similarity or visual plausibility, but by whether synthetic data behave plausibly under trading evaluation. Prior studies largely overlooked this criterion, leading to unstable outcomes. By embedding domain knowledge directly into training, SFAG highlights a principled path toward synthetic data that are both scientifically faithful and practically useful.

\textbf{Future work.}
This work opens several directions: (i) extending evaluation beyond a single equity index to multi-asset and cross-market settings, including FX, rates, and commodities, (ii) incorporating richer stylized facts such as tail asymmetry, volatility-of-volatility, and regime persistence, and (iii) exploring backbone-agnostic alignment, applying SFAG principles to diffusion and transformer-based generators for scalable and flexible financial simulation.

% Below is an example of how to insert images. Delete the ``\vspace'' line,
% uncomment the preceding line ``\centerline...'' and replace ``imageX.ps''
% with a suitable PostScript file name.
% -------------------------------------------------------------------------

%\vfill
%\pagebreak

\bibliographystyle{IEEEbib}
\bibliography{refs}

@article{cont2001empirical,
  title={Empirical properties of asset returns: stylized facts and statistical issues},
  author={Cont, Rama},
  journal={Quantitative finance},
  volume={1},
  number={2},
  pages={223},
  year={2001},
  publisher={IOP Publishing}
}

@article{wiese2020quant,
  title={Quant GANs: deep generation of financial time series},
  author={Wiese, Magnus and Knobloch, Robert and Korn, Ralf and Kretschmer, Peter},
  journal={Quantitative Finance},
  volume={20},
  number={9},
  pages={1419--1440},
  year={2020},
  publisher={Taylor \& Francis}
}

@article{mandelbrot1963variation,
 ISSN = {00219398, 15375374},
 URL = {http://www.jstor.org/stable/2350970},
 author = {Benoit Mandelbrot},
 journal = {The Journal of Business},
 number = {4},
 pages = {394--419},
 publisher = {University of Chicago Press},
 title = {The Variation of Certain Speculative Prices},
 urldate = {2026-01-19},
 volume = {36},
 year = {1963}
}

@article{engle1982autoregressive,
 ISSN = {00129682, 14680262},
 URL = {http://www.jstor.org/stable/1912773},
 author = {Robert F. Engle},
 journal = {Econometrica},
 number = {4},
 pages = {987--1007},
 publisher = {[Wiley, Econometric Society]},
 title = {Autoregressive Conditional Heteroscedasticity with Estimates of the Variance of United Kingdom Inflation},
 urldate = {2026-01-19},
 volume = {50},
 year = {1982}
}

@article{ding1993long,
  title={A long memory property of stock market returns and a new model},
  author={Ding, Zhuanxin and Granger, Clive WJ and Engle, Robert F},
  journal={Journal of empirical finance},
  volume={1},
  number={1},
  pages={83--106},
  year={1993},
  publisher={Elsevier}
}

@inproceedings{black1976studies,
  title={Studies of stock price volatility changes},
  author={Black, Fischer},
  booktitle={Proceedings of the 1976 Meeting of the Business and Economic Statistics Section},
  pages={177--181},
  year={1976}
}

@article{bouchaud2001leverage,
  title={Leverage effect in financial markets: The retarded volatility model},
  author={Bouchaud, Jean-Philippe and Matacz, Andrew and Potters, Marc},
  journal={Physical review letters},
  volume={87},
  number={22},
  pages={228701},
  year={2001},
  publisher={APS}
}

@article{calvet2002multifractality,
  title={Multifractality in asset returns: theory and evidence},
  author={Calvet, Laurent and Fisher, Adlai},
  journal={Review of Economics and Statistics},
  volume={84},
  number={3},
  pages={381--406},
  year={2002},
  publisher={MIT Press 238 Main St., Suite 500, Cambridge, MA 02142-1046, USA journals~…}
}

@inproceedings{bai2018tcn,
  title={An empirical evaluation of generic convolutional and recurrent networks for sequence modeling},
  author={Bai, Shaojie and Kolter, J Zico and Koltun, Vladlen},
  booktitle={Proceedings of the 35th International Conference on Machine Learning (ICML)},
  pages={481--490},
  year={2018}
}

@article{li2023gpttst,
  title={GPT-TST: Generative pre-trained transformer for time series forecasting},
  author={Li, Yan and Bai, Hongliang and Song, Xiaoyang and Cheng, Yu and Zhang, Xinyu},
  journal={Expert Systems with Applications},
  volume={230},
  pages={120675},
  year={2023},
  publisher={Elsevier}
}

@inproceedings{lim2021temporalfusion,
  title={Temporal Fusion Transformers for interpretable multi-horizon time series forecasting},
  author={Lim, Bryan and Arik, Sercan O and Loeff, Nicolas and Pfister, Tomas},
  booktitle={Proceedings of the 28th ACM SIGKDD Conference on Knowledge Discovery and Data Mining (KDD)},
  pages={2967--2975},
  year={2021}
}

@article{zhou2022fedformer,
  title={FEDformer: Frequency enhanced decomposed transformer for long-term series forecasting},
  author={Zhou, Haixu and Zhang, Shanghang and Peng, Jieqi and Zhang, Shuai and Li, Jianxin and Xiong, Hui and Zhang, Wancai},
  journal={Proceedings of the 39th International Conference on Machine Learning (ICML)},
  pages={27268--27286},
  year={2022}
}

@article{liu2025timeformer,
  title={TimeFormer: Transformer with Attention Modulation Empowered by Temporal Characteristics for Time Series Forecasting},
  author={Liu, Zhipeng and Duan, Peibo and Tang, Xuan and Li, Baixin and Huang, Yongsheng and Geng, Mingyang and Zhang, Changsheng and Zhang, Bin and Wang, Binwu},
  journal={Expert Systems with Applications},
  pages={131040},
  year={2025},
  publisher={Elsevier}
}

@inproceedings{rasul2021timegrad,
  title={Autoregressive Denoising Diffusion Models for Multivariate Probabilistic Time Series Forecasting},
  author={Kashif Rasul and Calvin Seward and Ingmar Schuster and Roland Vollgraf},
  booktitle={Proceedings of the 38th International Conference on Machine Learning (ICML)},
  pages  = {8857--8868},
  year={2021},
  url={https://api.semanticscholar.org/CorpusID:231719657}
}

@inproceedings{song2021score,
  author       = {Yang Song and
                  Jascha Sohl{-}Dickstein and
                  Diederik P. Kingma and
                  Abhishek Kumar and
                  Stefano Ermon and
                  Ben Poole},
  title        = {Score-Based Generative Modeling through Stochastic Differential Equations},
  booktitle    = {9th International Conference on Learning Representations (ICLR)},
  year         = {2021}
}

@article{ericson2024deep,
  title={Deep generative modeling for financial time series with application in VaR: A comparative review},
  author={Ericson, Lars and Zhu, Xuejun and Han, Xusi and Fu, Rao and Li, Shuang and Guo, Steve and Hu, Ping},
  journal={arXiv preprint arXiv:2401.10370},
  year={2024}
}

@article{rizzato2023generative,
title = {Generative Adversarial Networks applied to synthetic financial scenarios generation},
journal = {Physica A: Statistical Mechanics and its Applications},
volume = {623},
pages = {128899},
year = {2023},
issn = {0378-4371},
doi = {https://doi.org/10.1016/j.physa.2023.128899},
url = {https://www.sciencedirect.com/science/article/pii/S0378437123004545},
author = {Matteo Rizzato and Julien Wallart and Christophe Geissler and Nicolas Morizet and Noureddine Boumlaik}
}

@article{dechant2025quantum,
	author={Dechant, David and Schwander, Eliot Jan Etienne and Van Drooge, Lucas and Moussa, Charles and Garlaschelli, Diego and Dunjko, Vedran and Tura Brugués, Jordi},
	title={Quantum generative modeling for financial time series with temporal correlations},
	journal={Machine Learning: Science and Technology},
	url={http://iopscience.iop.org/article/10.1088/2632-2153/ae39a2},
	year={2026},
}

@inproceedings{wgangp2017,
author = {Gulrajani, Ishaan and Ahmed, Faruk and Arjovsky, Martin and Dumoulin, Vincent and Courville, Aaron},
title = {Improved training of wasserstein GANs},
year = {2017},
booktitle = {Proceedings of the 31st International Conference on Neural Information Processing Systems (NIPS)},
pages = {5769–5779},
numpages = {11}
}

@inproceedings{shen2018wasserstein,
author = {Shen, Jian and Qu, Yanru and Zhang, Weinan and Yu, Yong},
title = {Wasserstein distance guided representation learning for domain adaptation},
year = {2018},
isbn = {978-1-57735-800-8},
booktitle = {Proceedings of the Thirty-Second AAAI Conference on Artificial Intelligence},
articleno = {497},
pages     = {4058--4065},
numpages = {8},
location = {New Orleans, Louisiana, USA},
}

\end{document}